# POLARIZATION PHENOMENA IN PROCESS $e^+ + e^- \to N + \bar{N} + \gamma$

G.I. Gakh[1,2], M.I. Konchatnij[1,2], N.P. Merenkov[1,2] and A. G. Gakh[2]

[1]*NSC "Kharkov Institute of Physics and Technology"*
*Akademicheskaya, 1, 61108 Kharkov, Ukraine*
[2]*V.N. Karazin Kharkiv National University, 61022 Kharkov, Ukraine*
e-mail: konchatnij@kipt.kharkov.ua

The emission of the hard photon from the initial state is considered. The nucleon polarization and the differential cross sections for some experimental conditions have been calculated. The case of the emission of the collinear (with respect to the direction of the electron beam momentum) photon is considered separately. The differential cross section, the nucleon polarization, the correlation coefficients for both polarized nucleons (provided the electron beam is unpolarized or longitudinally polarized), the transfer polarization from the longitudinally polarized electron beam to the nucleon have been calculated. The photon energy distribution for the reaction $e^+ e^- \to h_1 h_2 \gamma$, where $h_1$ and $h_2$ are some hadrons for the case of the collinear photon, emitted in the initial state, has been calculated. As $h_1 h_2$ final state we considered some channels, namely: two spinless mesons (for example, $\pi^+\pi^-, K^+K^-$), two spin–one particles (for example, $\rho^+\rho^-, d\bar{d}$), and the channels $\pi a_1(1260)$ and $\Delta(1232)\bar{N}$. The photon energy distributions are calculated in terms of the form factors of the $\gamma^* \to h_1 h_2$ transition ($\gamma^*$ is the virtual photon).

## INTRODUCTION

The investigation of the electromagnetic form factors of the proton and neutron in both the space–like and time–like regions of the momentum transfer squared is important for the understanding of the internal structure of these particles and for the interpretation of many data on reactions with participation of the nucleons. The knowledge of the nucleon form factors is also required for the interpretation of the nuclear structure and various measurements of the reactions involving nuclei. So, the experimental determination of the elastic nucleon electromagnetic form factors in the region of small and large momentum transfer squared is one of the major fields of research in hadron physics (see the review [1]).

The measurement of the nucleon electromagnetic form factors in the space–like region of the momentum transfer squared has a long history. The electric and magnetic form factors were determined both for the proton and neutron using two different techniques: the Rosenbluth separation [2, 3] and polarization transfer method [4, 5, 6, 7]. It turned out that the measurements of the ratio of the proton electric and magnetic form factors using these two methods lead to the appreciably different results, and this difference is increasing when $Q^2$ (the four– momentum transfer squared) grows. The ratio $G_E^p / G_M^p$ is monotonically decreasing with increasing $Q^2$ suggesting crossing zero at $Q^2 \approx (8-9) \; GeV^2$ [8]. These unexpected results revived an experimental and theoretical investigations of this problem (see reviews [1, 9]). One possible mechanism suggested for the explanation of this discrepancy is the two–photon–exchange contribution to the elastic electron–nucleon scattering [10, 11]. Other considerations lead to the conclusion that the contribution from the two–photon term is too small at the $Q^2$ values of interest [12] and/or lead to a definite non–linearity in the Rosenbluth plot which has not been seen in the data so far [13]. A model independent study of the two–photon–exchange mechanism in the elastic electron–nucleon scattering and its consequences on the experimental observables has been carried on in Refs. [14, 15, 16], and in the crossed channels: proton–antiproton annihilation into the lepton pair [17] and annihilation of the electron–positron pair into the nucleon–antinucleon [18].

The data on nucleon form factors in the time–like region are not numerous. So, the separation the electric and magnetic form factors in this region has not yet been done. One of the reasons is the limitation in the intensity of antiproton beams and of the luminosity of electron-positron colliders.

Nevertheless, a few unexpected results have been observed in the measurements of the nucleon form factors in the time–like region (note that the accuracy of the data set is not sufficiently good to do definite statements). Despite of the relation $|G_E| = |G_M|$, which must be valid at the threshold of the $e^+e^- \to N\bar{N}$ reaction, the neutron electric form factor is negligible near the threshold as may be suggested from the measurement of the differential cross section. The general behavior of the neutron time–like form factors is rather unexpected. The proton magnetic form factor is smaller than the neutron one at $q^2 \leq 6 GeV^2$ (where experiments were done). The review of the present status in this field of investigations is given in [1]. Note also that in the time–like region the proton magnetic form factor is considerably bigger than the corresponding space–like quantity.

Recent experimental data on the nucleon form factors (both in the space– and time–like regions) together with new theoretical developments [19] (where the analytic continuation of the QCD formulas from the space–like region of momentum transfer to the time–like one was discussed) show the necessity of a global description of the nucleon form factors in the full region of the $q^2$ variable. Some papers were already appeared [20, 21, 22].

The experimental data on the time–like form factors may turn out to be very sensitive to the details of existing models. For example, the analysis, performed in Ref. [23], taking into account the combined space–like and time–like data on the proton and neutron form factors leads to a good fit to the space–like form factors but cannot describe neutron time–like data.

So, the experimental investigation of the nucleon form factors in the time–like region may give additional valuable information about the internal nucleon structure and can test the existing models.

In the time–like region, the nucleon form factors can be measured using the reactions $e^+e^- \to N\bar{N}$ or $\bar{p}p \to e^+e^-$. In this region only a small set of data exists. The neutron form factors were measured by the FENICE collaboration [24], using the ADONE $e^+e^-$ collider in Frascati. The reaction $e^+e^- \to n\bar{n}$ was studied up to $q^2 \approx 6\,GeV^2$ starting with reaction threshold. The proton form factors were measured in a broader region of $q^2$ values. The region of the large $q^2$ was achieved with the help of $\bar{p}p \to e^+e^-$ reaction: Fermilab experiment E835 measured the cross section of this reaction up to $q^2 = 18.22\,GeV^2$ [25].

Some experiments are planning to study this region of $q^2$. A new experiment at an asymmetric collider is proposed at SLAC with the ambitious goal to measure the nucleon form factors from threshold up to 3 GeV with the same accuracy currently available in the space–like region [26].

The chance of measuring these form factors with higher precision will be given by a suitable upgrade of *DAΦNE* energy [27]. The number of good detectable events per day is about, or exceeding, the total amount of events collected by *FENICE* in all its data takings.

*FINUDA* planned to offer a unique possibility - a measurement of the nucleon polarization [27]. This kind of measurement would be of great interest, as it would be a handle to infer something about the relative phases of $G_M$ and $G_E$ form factors.

As it is known, the $e^+e^- \to N\bar{N}$ reaction is the cross channel for the reaction of the elastic electron–nucleon scattering. The form factors describing the annihilation channel are assumed to be the analytical continuation of the space–like ones. So, one may expect that the problems existed in the scattering channel will also manifest itself in the annihilation channel. It concerns, in particular, the problem of the two–photon–exchange contribution.

Theoretically, the reaction $e^+e^- \to N\bar{N}$ was studied in a number of papers. The dependence of polarization states of created one–half spin baryons in the $e^+e^- \to B\bar{B}$ reaction on the polarization of colliding $e^+e^-$ - beams was investigated in Ref. [28]. The formulae obtained in this paper exhaust all polarization effects of baryons with spin 1/2 in the $e^+e^- \to B\bar{B}$ reaction. Numerical estimates of polarization effects were presented only for the nucleons. The polarization effects appear to be very sensitive to the choice of the nucleon form factors parametrization and are rather large in absolute value. The pronounced energy dependence measured in the cross section of the reactions $e^+e^- \to p\bar{p}$ investigated in Ref. [29] in the near-threshold region. The authors considered the role of the antinucleon-nucleon interaction in the initial- or final state using $N\bar{N}$ potential derived within chiral effective field theory.

The existence of the T–odd single–spin asymmetry normal to the reaction scattering plane requires a non–zero phase difference between the electric and magnetic form factors. The measurement of the polarization of one of the outgoing nucleons allows to determine the phase of the ratio $G_E / G_M$. In Ref. [30] it was shown that measurements of the proton polarization in $e^+e^- \to p\bar{p}$ reaction strongly discriminate between the analytic forms of models suggested to fit the proton data in the space–like region.

As it is known, the problem of taking into account the radiative corrections in the elastic electron–nucleon scattering is important for the reliable extraction of the nucleon form factors. The same is valid for the crossed channel. The importance of the $e^- + e^+ \to N + \bar{N} + \gamma$ reaction is not only due to the fact that it is a part of the radiative corrections to the $e^- + e^+ \to N + \bar{N}$ reaction but rather because it allows to measure the nucleon form factors by the radiative return method [31].

In this paper we investigate the polarization phenomena in the reaction

$$e^-(k_1) + e^+(k_2) \to N(p_1) + \bar{N}(p_2) + \gamma(k), \tag{1}$$

where four–momenta of the corresponding particles are given in the brackets. We consider here the emission of the additional hard photon by the initial electron or positron since the emission of the photon by the final state particles is model dependent and suppressed with respect to the initial state radiation due to the large nucleon mass as compared with electron one and perhaps by the nucleon form factors.

Here we derive the expressions for the differential cross section and various polarization observables taking into account the nucleon form factors.

We consider a particular case of the high–energy photon emission at small angles (the radiative return). The differential cross section and various polarization observables (the nucleon polarization, the correlation coefficients for the nucleon–antinucleon pair and polarization transfer from the longitudinally polarized electron to the nucleon), when the angular distribution of the nucleon and energy of the emitted photon are measured, have been calculated for the case of the photon emitted at small angles relative to the electron beam momentum.

The standard analysis of the experimental data requires the account for all possible systematic uncertainties. One of the important source of such uncertainties are the electromagnetic radiative effects caused by physical processes which take place in higher orders of the perturbation theory with respect to the electromagnetic interaction. In present paper we calculate the model–independent QED radiative corrections to the observables (both polarized and unpolarized). Our approach is based on the covariant parametrization of the nucleon or antinucleon spin four-vectors in terms of the four–momenta of the particles in process (1) [32, 33].

The photon energy distribution for the reaction $e^+e^- \to h_1 h_2 \gamma$, where $h_1$ and $h_2$ are some hadrons for the case of the collinear photon, emitted in the initial state, has been calculated. As $h_1 h_2$ final state we considered some channels, namely: two spinless mesons (for example, $\pi^+\pi^-, K^+K^-$), two spin–one particles (for example, $\rho^+\rho^-, d\bar{d}$), the $\pi a_1(1260)$ and $\Delta(1232)\bar{N}$ channels. The photon energy distributions are calculated in terms of the form factors of the $\gamma^* \to h_1 h_2$ transition ($\gamma^*$ is the virtual photon).

The paper is organized as follows. In Section "POLARIZATION PHENOMENA IN $e^+ + e^- \to N + \bar{N}$" the polarization phenomena for the Born approximation, i.e., for the reaction $e^- + e^+ \to N + \bar{N}$ are given. In Section "HARD–PHOTON EMISSION" the emission of the hard photon by the initial state is considered. The nucleon polarization and the differential cross sections for some experimental conditions have been calculated. In Section "RADIATIVE RETURN. SMALL ANGLES" the emission of the collinear photon is considered in details. The differential cross section and various polarization observables have been calculated. In Section "PHOTON ENERGY DISTRIBUTION" we have calculated the photon energy distribution for the reaction $e^+e^- \to h_1 h_2 \gamma$, where $h_1$ and $h_2$ are some hadrons for the case of the collinear photon, emitted in the initial state. As $h_1 h_2$ final state we consider some channels. In Section "RADIATIVE RETURN. LARGE ANGLES" the emission of the collinear photon at large angles is considered. The main results are summarized in Section "CONCLUSIONS".

## POLARIZATION PHENOMENA IN $e^+ + e^- \to N + \bar{N}$

Let us consider first the production of $N\bar{N}$-pair without emission of additional photons:

$$e^-(k_1) + e^+(k_2) \to N(p_1) + \bar{N}(p_2), \tag{2}$$

where four–momenta of the corresponding particles are given in the brackets. The matrix element of this reaction can be written as follows

$$M = i4\pi \frac{\alpha}{q^2} j_\mu J_\mu, \tag{3}$$

where $\alpha = e^2/4\pi = 1/137$, $q = k_1 + k_2 = p_1 + p_2$ is the virtual photon four-momentum. The leptonic and hadronic currents can be written as

$$j_\mu = \bar{v}(k_2) \gamma_\mu u(k_1), \tag{4}$$

$$J_\mu = \bar{u}(p_1)[F_1(q^2)\gamma_\mu - \frac{1}{2M} F_2(q^2) \sigma_{\mu\nu} q_\nu] v(p_2),$$

where $\sigma_{\mu\nu} = (\gamma_\mu \gamma_\nu - \gamma_\nu \gamma_\mu)/2$, $M$ is the nucleon mass and $F_1(q^2)$ and $F_2(q^2)$ are the Dirac and Pauli nucleon electromagnetic form factors, respectively, which are the complex functions of the variable $q^2$. The complexity of the form factors arises due to the final–state strong interaction of the produced $N\bar{N}$ − pair. In the following, we use the standard magnetic $G_M(q^2)$ and charge $G_E(q^2)$ nucleon form factors which are related to the form factors $F_1(q^2)$ and

$F_2(q^2)$ as follows

$$G_M = F_1 + F_2, G_E = F_1 + \tau F_2, \tau = \frac{q^2}{4M^2}. \tag{5}$$

Then, the differential cross section of the $e^- + e^+ \to N + \bar{N}$ reaction, for the case of the polarized electron beam and unpolarized positron beam, can be written as follows in the reaction centre of mass system (CMS)

$$\frac{d\sigma}{d\Omega} = \frac{\alpha^2}{8q^6} \beta L_{\mu\nu} H_{\mu\nu}, \tag{6}$$

where $\beta$ is the nucleon velocity in CMS, $\beta = \sqrt{1 - 4M^2/q^2}$, and the leptonic and hadronic tensors are defined as

$$L_{\mu\nu} = j_\mu j_\nu^*, H_{\mu\nu} = J_\mu J_\nu^*.$$

The leptonic tensor for the case of longitudinally polarized electron beam has the form (other components of the electron polarization lead to the observables suppressed by a factor $m/M$, where $m$ is the electron mass)

$$L_{\mu\nu} = -q^2 g_{\mu\nu} + 2(k_{1\mu} k_{2\nu} + k_{1\nu} k_{2\mu}) + 2i\lambda_e <\mu\nu q k_2>, \tag{7}$$

where $<\mu\nu ab> = \varepsilon_{\mu\nu\rho\sigma} a_\rho b_\sigma$ and $\lambda_e$ is the degree of the electron longitudinal polarization (we use the following definition for the antisymmetric tensor $\varepsilon_{1230} = 1$).

Taking into account the polarization states of the produced nucleon and antinucleon, the hadronic tensor can be written as a sum of four contributions as follows:

$$H_{\mu\nu} = \frac{1}{4} H_{\mu\nu}(0) + \frac{1}{2} H_{\mu\nu}(1) + \frac{1}{2} \bar{H}_{\mu\nu}(1) + H_{\mu\nu}(2), \tag{8}$$

where the tensor $H_{\mu\nu}(0)$ describes the production of unpolarized nucleon and antinucleon, the tensor $H_{\mu\nu}(1)(\bar{H}_{\mu\nu}(1))$ describes the production of polarized nucleon (antinucleon) and the tensor $H_{\mu\nu}(2)$ corresponds to the production of polarized particles, nucleon and antinucleon.

Let us consider the production of unpolarized $N\bar{N}-$ pair as a result of annihilation of unpolarized $e^+ e^- -$ pair. In this case the general structure of the hadronic tensor can be written as

$$H_{\mu\nu}(0) = H_1 \tilde{g}_{\mu\nu} + \frac{1}{M^2} H_2 \tilde{p}_{1\mu} \tilde{p}_{1\nu}, \tag{9}$$

where $\tilde{g}_{\mu\nu} = g_{\mu\nu} - q_\mu q_\nu / q^2$ and $\tilde{p}_{1\mu} = p_{1\mu} - q \cdot p_1 q_\mu / q^2$. One can get the following expressions for these structure functions for the case of the hadronic current given by Eq. (4)

$$H_1(q^2) = -2q^2 |G_M(q^2)|^2, H_2(q^2) = \frac{8M^2}{1-\tau}[\tau |G_M(q^2)|^2 - |G_E(q^2)|^2]. \tag{10}$$

Then, the contraction of the leptonic and hadronic tensors, in the case of unpolarized initial beams and produced nucleon and antinucleon, can be written as

$$S(0) = L_{\mu\nu} H_{\mu\nu}(0) = -2q^2 H_1(q^2) + H_2(q^2)[(\tau-1)q^2 - \frac{(u-t)^2}{4M^2}],$$

where $u = (k_1 - p_2)^2$ and $t = (k_1 - p_1)^2$.

The differential cross section of the $e^- + e^+ \to N + \bar{N}$ reaction, for the case of unpolarized particles, has the form in CMS

$$\frac{d\sigma_{un}}{d\Omega} = \frac{\alpha^2 \beta}{4q^2} D, \tag{11}$$

$$D = (1+cos^2\theta)|G_M(q^2)|^2 + \frac{1}{\tau}sin^2\theta|G_E(q^2)|^2,$$

where $\theta$ is the angle between the electron and detected nucleon momenta in the $e^+ + e^- \to N + \bar{N}$ reaction CMS. This expression coincides with the result for the differential cross section obtained in Ref. [28]. At the threshold of the reaction, $q^2 = 4M^2$, we have $G_M = G_E = G$ (this relation follows from the definition (5)) and the formula (11) reduced to (near the threshold)

$$\frac{d\sigma^{th}_{un}}{d\Omega} = \frac{\alpha^2 \beta}{2q^2} D^{th}, D^{th} = |G|^2. \tag{12}$$

Integrating the expression (11) over the angular variables, we obtain the total cross section for the reaction $e^+ + e^- \to N + \bar{N}$

$$\sigma_{un}(q^2) = \frac{4\pi}{3}\frac{\alpha^2 \beta}{q^2}[|G_M(q^2)|^2 + \frac{2M^2}{q^2}|G_E(q^2)|^2]. \tag{13}$$

Now, let us consider the single polarization observables. To do this, it is necessary to calculate the hadronic tensor for the case when produced nucleon is polarized. We can write this tensor as a sum of two terms: one is symmetrical and another one is antisymmetrical (over $\mu$ and $\nu$ indices)

$$H_{\mu\nu}(1) = S_{\mu\nu} + iA_{\mu\nu},$$

$$S_{\mu\nu} = \frac{2}{M}\frac{1}{1-\tau} Im G_E G_M^* (\tilde{p}_{1\mu} a_\nu + \tilde{p}_{1\nu} a_\mu), \tag{14}$$

$$A_{\mu\nu} = 2M Re G_E G_M^* <\mu\nu q s_1> + \frac{1}{M}\frac{q \cdot s_1}{1-\tau}[|G_M|^2 - Re G_E G_M^*]<\mu\nu p_1 q>,$$

where $a_\mu = \varepsilon_{\mu\nu\lambda\sigma} p_{1\nu} p_{2\lambda} s_{1\sigma}$ and $s_{1\mu}$ is the nucleon polarization four–vector and $s_1 \cdot p_1 = 0$, $s_1^2 = -1$.

Let us define the coordinate frame in CMS of the $e^- + e^+ \to N + \bar{N}$ reaction. The $z$ axis is directed along the momentum of the nucleon ($\vec{p}$), $y$ axis is orthogonal to the reaction plane and directed along the vector $\vec{k} \times \vec{p}$, where $\vec{k}$ is the electron momentum, and the $x$ axis forms a left–handed coordinate system. Therefore, the components of the unit vectors are: $\hat{\vec{p}} = (0,0,1)$ and $\hat{\vec{k}} = (-sin\theta, 0, cos\theta)$ with $\hat{\vec{p}} \cdot \hat{\vec{k}} = cos\theta$.

The polarization 4–vector $s_{1\mu}$ of a nucleon in the system where it has momentum $\vec{p}$ is connected with the polarization vector $\vec{\chi}_1$ in its rest frame by a Lorentz boost

$$\vec{s}_1 = \vec{\chi}_1 + \frac{\vec{p} \cdot \vec{\chi}_1 \vec{p}}{M(E+M)}, s_{10} = \frac{1}{M}\vec{p} \cdot \vec{\chi}_1. \tag{15}$$

Let us note that four-vector $s_{1\mu}$ can be written down as

$$s_{1\mu} = \sum s_{1\mu}^{(i)} \chi_{1i}, \quad i = x, y, z.$$

Each four-vector $s_{1\mu}^{(i)}$ satisfies the following conditions

$$s_1^{(i)} \cdot p_1 = 0, s_1^{(i)2} = -1.$$

Note that the polarization 4–vectors of the particles can be parameterized in terms of the four–momenta of these particles in the reaction under study (it is very convenient when calculating the radiative corrections to this reaction).

Let us write the chosen axes in a covariant form in terms of the four–momenta. So, in the reaction CMS we choose the longitudinal direction $\vec{l}$ ($z$ axis) along the nucleon momentum and the transverse one $\vec{t}$ in the plane $(\vec{p},\vec{k})$ ($x$ axis) and perpendicular to $\vec{l}$ ($y$ axis), then

$$P_\mu^{(l)} = \frac{1}{d_1}(q^2 p_{1\mu} - 2M^2 q_\mu), d_1^2 = M^2 q^2(q^2 - 4M^2),$$

$$P_\mu^{(t)} = \frac{1}{d_2}[(q^2 - 4p_1\cdot k_1)p_{1\mu} + 2(p_1\cdot k_1 - M^2)q_\mu + (4M^2 - q^2)k_{1\mu}],$$

$$d_2^2 = (q^2 - 4M^2)(4p_1\cdot k_1 p_1\cdot k_2 - M^2 q^2),$$

$$P_\mu^{(n)} = \frac{1}{d_3}<\mu q k_1 p_1>, d_3^2 = \frac{q^2}{4}(4p_1\cdot k_1 p_1\cdot k_2 - M^2 q^2). \tag{16}$$

It can be verified that the set of the four–vectors $P_\mu^{(l,t,n)}$ has the properties

$$P_\mu^{(\alpha)} P_\mu^{(\beta)} = -\delta_{\alpha\beta}, P_\mu^{(\alpha)}\cdot p_{1\mu} = 0, \alpha, \beta = l,t,n \tag{17}$$

and that in the reaction CMS we have

$$P_\mu^{(l)} = \frac{1}{M}(|\vec{p}|, E\vec{n}_1), P_\mu^{(t)} = (0,\vec{t}), P_\mu^{(n)} = (0,\vec{n}), \tag{18}$$

$$\vec{n}_1 = \frac{\vec{p}}{|\vec{p}|}, \vec{n}_2 = \frac{\vec{k}}{|\vec{k}|}, \vec{t} = \frac{(\vec{n}_1\cdot\vec{n}_2)\vec{n}_1 - \vec{n}_2}{\sqrt{1 - (\vec{n}_1\cdot\vec{n}_2)^2}}, \vec{n} = \frac{\vec{n}_2\times\vec{n}_1}{\sqrt{1-(\vec{n}_1\cdot\vec{n}_2)^2}}.$$

It is easy to show that the following relations are valid

$$P_\mu^{(l)} = s_{1\mu}^{(z)}, P_\mu^{(t)} = s_{1\mu}^{(x)}, P_\mu^{(n)} = s_{1\mu}^{(y)}. \tag{19}$$

Note that, unlike the elastic electron–nucleon scattering in the Born approximation, the hadronic tensor $H_{\mu\nu}$ (1) in the time–like region contains the symmetric part even in the Born approximation due to the complexity of the nucleon form factors. So, this term leads to the non–zero polarization of the outgoing nucleon (the initial state is unpolarized) in the $e^- + e^+ \to N + \bar{N}$ reaction and it can be written as

$$P_y = \frac{sin2\theta}{\sqrt{\tau}D} Im G_M G_E^*. \tag{20}$$

This expression gives the well known result for the polarization $P_y$ obtained in Ref. [28]. One can see also that:
- The polarization of the outgoing nucleon, in this case, is determined by the polarization component which is perpendicular to the reaction plane.
- The polarization, being T–odd quantity, does not vanish even in the one–photon–exchange approximation due to the complexity of the nucleon form factors in the time–like region (to say more exactly, due to the non–zero difference of the phases of these form factors). This is principal difference with the elastic electron–nucleon scattering.
- In the Born approximation this polarization becomes equal to zero at the scattering angle $\theta = 90^0$ (as well as at $\theta = 0^0$ and $180^0$).

In the threshold region we can conclude that in the Born approximation this polarization must be zero due to the relation $G_E = G_M$ which is valid at the threshold.

If one of the colliding beam is longitudinally polarized then nucleon acquires $x-$ and $z-$ components of the polarization, which lie in the $e^+ + e^- \to N + \bar{N}$ reaction plane. These components can be written as (we assume 100% polarization of the electron beam)

$$P_x = -\frac{2\sin\theta}{\sqrt{\tau}D}ReG_M G_E^*, P_z = \frac{2}{D}\cos\theta |G_M|^2. \tag{21}$$

These polarization components are T–even observables and they are non–zero in the Born approximation even for the elastic electron–nucleon scattering. Note that in the Born approximation we obtain the result of Ref. [28]. The polarization component $P_z$ equals to zero at the scattering angle $\theta = 90^0$ in the Born approximation. Transversally polarized electron beam leads to the nucleon polarization which is smaller by factor $(m/M)$ than for the case of the longitudinal polarization of the electron beam.

Let us consider the case when the produced antinucleon and nucleon are both polarized. The corresponding hadronic tensor can be written as

$$H_{\mu\nu}(2) = S_{\mu\nu}(s_1, s_2) + iA_{\mu\nu}(s_1, s_2),$$

$$A_{\mu\nu}(s_1, s_2) = [q \cdot s_1(\tilde{s}_{2\mu}\tilde{p}_{1\nu} - \tilde{s}_{2\nu}\tilde{p}_{1\mu}) - q \cdot s_2(\tilde{s}_{1\mu}\tilde{p}_{1\nu} - \tilde{s}_{1\nu}\tilde{p}_{1\mu})]\frac{ImG_E G_M^*}{\tau - 1},$$

$$S_{\mu\nu}(s_1, s_2) = A_g \tilde{g}_{\mu\nu} + A_{11}\tilde{p}_{1\mu}\tilde{p}_{1\nu} + A_s(\tilde{s}_{1\mu}\tilde{s}_{2\nu} + \tilde{s}_{1\nu}\tilde{s}_{2\mu}) + A_{1s}[q \cdot s_1(\tilde{p}_{1\mu}\tilde{s}_{2\nu} + \tilde{p}_{1\nu}\tilde{s}_{2\mu}) - q \cdot s_2(\tilde{p}_{1\mu}\tilde{s}_{1\nu} + \tilde{p}_{1\nu}\tilde{s}_{1\mu})], \tag{22}$$

$$\tilde{s}_{i\mu} = s_{i\mu} - \frac{q \cdot s_i}{q^2}q_\mu, \quad i = 1, 2,$$

where $s_{2\mu}$ is the antinucleon polarization 4-vector ($p_2 \cdot s_2 = 0$). Antisymmetric part of the tensor $H_{\mu\nu}(2)$ arises due to the fact that nucleon form factors in the time–like region are complex quantities.

The structure functions $A_i$ have the following form

$$A_g = \frac{1}{2}(q^2 s_1 \cdot s_2 - 2q \cdot s_1 q \cdot s_2)|G_M|^2,$$

$$A_{11} = 2\frac{s_1 \cdot s_2}{\tau - 1}[\tau|G_M|^2 - |G_E|^2] + \frac{q \cdot s_1 q \cdot s_2}{M^2(\tau-1)^2}|G_E - G_M|^2,$$

$$A_s = -\frac{q^2}{2}|G_M|^2, A_{1s} = \frac{1}{\tau - 1}Re(\tau G_M - G_E)G_M^*. \tag{23}$$

Using previous formulae one can obtain the following expressions for the components of the polarization correlation tensor $P_{ik}(i, k = x, y, z)$ of the nucleon and antinucleon, created by the one–photon–exchange mechanism in the $e^+ + e^- \to N + \bar{N}$ process:

$$P_{xx} = \frac{\sin^2\theta}{\tau D}[\tau|G_M|^2 + |G_E|^2], P_{yy} = \frac{\sin^2\theta}{\tau D}[|G_E|^2 - \tau|G_M|^2],$$

$$P_{zz} = \frac{1}{\tau D}[\tau(1 + \cos^2\theta)|G_M|^2 - \sin^2\theta|G_E|^2],$$

$$P_{xz} = P_{zx} = -2\frac{\sin\theta}{\sqrt{\tau}D}\cos\theta(ReG_M G_E^*), \tag{24}$$

where the first index of the tensor $P_{ik}$ refers to the component of the nucleon polarization vector, whereas the second index refers to the component of the antinucleon polarization vector.

The antinucleon polarization four-vector, $s_{2\mu}$, is described by the formula (15) where it is necessary to do the following substitution: $\vec{p} \to -\vec{p}$ and $\vec{\chi}_1 \to \vec{\chi}_2$ ($\vec{\chi}_2$ is the polarization vector of the antinucleon in its rest frame). The antinucleon polarization 4–vectors $\overline{P}_\mu^{(i)}, i = l, t, n$ (in terms of the particles four–momenta) can be written down as

$$\overline{P}_\mu^{(l)} = -\frac{1}{d_1}(q^2 p_{2\mu} - 2M^2 q_\mu), \overline{P}_\mu^{(n)} = \frac{1}{d_3} <\mu q k_2 p_2>, \qquad (25)$$

$$\overline{P}_\mu^{(t)} = \frac{1}{d_2}[(q^2 - 4p_2 \cdot k_1)p_{2\mu} + 2(p_2 \cdot k_1 - M^2)q_\mu + (4M^2 - q^2)k_{1\mu}].$$

It is easy to show that the following relations are valid

$$\overline{P}_\mu^{(l)} = s_{2\mu}^{(z)}, \overline{P}_\mu^{(t)} = s_{2\mu}^{(x)}, \overline{P}_\mu^{(n)} = s_{2\mu}^{(y)}. \qquad (26)$$

And for the completeness we give here the non–zero coefficients for the case of the longitudinally polarized electron beam

$$P_{xy} = P_{yx} = 0, P_{zy} = P_{yz} = \frac{\sin\theta}{\sqrt{\tau}D} Im(G_M G_E^*). \qquad (27)$$

The following relation exists for these coefficients
$$P_{xx} + P_{yy} + P_{zz} = 1.$$

One can see that:
- The components of the tensor describing the polarization correlations $P_{xx}$, $P_{yy}$, $P_{zz}$, $P_{xz}$, and $P_{zx}$ are the T–even observables, whereas the components $P_{yz}$, and $P_{zy}$ are the T–odd ones.
- In the Born approximation the expressions for the T–odd polarization correlations coincide with the corresponding components of the polarization correlation tensor of baryon $B$ and antibaryon $\overline{B}$ created by the one–photon–exchange mechanism in the $e^+ + e^- \to B + \overline{B}$ process [28]. The expressions for the T–even polarization correlations calculated in this paper have some misprints.

### HARD–PHOTON EMISSION

Let us consider the emission of the hard photon, i.e., consider the reaction (1). In this case the matrix element corresponding to the emission of photon by electron and positron, can be written as

$$M^\gamma = i4\pi\alpha\sqrt{4\pi\alpha}\frac{1}{q^2} j_\mu^\gamma J_\mu, \qquad (28)$$

where the leptonic current with emission of additional photon has the form

$$j_\mu^\gamma = -\overline{v}(k_2)[\frac{1}{t_1}\gamma_\mu(\hat{k}_1 - \hat{k} + m)\hat{A} + \frac{1}{t_2}\hat{A}(\hat{k} - \hat{k}_2 + m)\gamma_\mu]u(k_1), \qquad (29)$$

where $t_1 = 2k \cdot k_1$, $t_2 = 2k \cdot k_2$, $A_\mu$ is the photon polarization four-vector and the square of the matrix element can be written as follows

$$|M^\gamma|^2 = 64\pi^3\alpha^3 \frac{1}{q^4} L_{\mu\nu}^\gamma H_{\mu\nu}, L_{\mu\nu}^\gamma = j_\mu^\gamma j_\nu^{\gamma*}, \qquad (30)$$

where the hadronic tensor has the same form as in the Born approximation but the structure functions defining this tensor depend on the shifted momentum transfer $q = k_1 + k_2 - k$.

Let us represent the leptonic tensor $L_{\mu\nu}^\gamma$ as a sum of the spin–independent and spin–dependent part (we consider only the case of the longitudinally polarized electron beam)

$$L^\gamma_{\mu\nu} = L^\gamma_{\mu\nu}(0) + L^\gamma_{\mu\nu}(s_e), \tag{31}$$

where the spin–independent part of this tensor can be written as

$$L^\gamma_{\mu\nu}(0) = A_0 \tilde{g}_{\mu\nu} + A_1 \tilde{k}_{1\mu}\tilde{k}_{1\nu} + A_2 \tilde{k}_{2\mu}\tilde{k}_{2\nu} + A_3(\tilde{k}_{1\mu}\tilde{k}_{2\nu} + \tilde{k}_{1\nu}\tilde{k}_{2\mu}), \tag{32}$$

where $\tilde{k}_{i\mu} = k_{i\mu} - (q \cdot k_i / q^2) q_\mu$, $(i = 1, 2)$ and the structure functions $A_i$ are

$$A_0 = 2(2m^2 q^2 - t_1 t_2)(\frac{1}{t_1^2} + \frac{1}{t_2^2}) - 8 k_1 \cdot k_2 \frac{q^2}{t_1 t_2}, A_1 = \frac{16m^2}{t_2^2} - \frac{8q^2}{t_1 t_2},$$

$$A_2 = \frac{16m^2}{t_1^2} - \frac{8q^2}{t_1 t_2}, A_3 = -\frac{16m^2}{t_1 t_2}. \tag{33}$$

The spin–independent part of the leptonic tensor $L^\gamma_{\mu\nu}(0)$ coincides with the one obtained in Ref. [34] and if we neglect the $m^2/t_1 t_2$ term in the structure functions $A_i$ then the result coincides with the expression obtained in Ref. [35].

Let us consider the spin–dependent part of the leptonic tensor $L^\gamma_{\mu\nu}(s_e)$. As it was noted above, we consider only the longitudinal polarization of the electron beam. The spin four–vector describing this polarization is $s_{e\mu} = \lambda_e k_{1\mu}/m_e$ in calculations of polarization effects in the Born approximation (this gives sufficient accuracy when we neglect the electron mass in the Born approximation). But it is necessary to use more correct expression for the spin four–vector. We use the following form

$$s_{e\mu} = \frac{1}{N}[(s - 2m^2)k_{1\mu} - 2m^2 k_{2\mu}], N^2 = m^2 s(s - 4m^2), \tag{34}$$

where $s = (k_1 + k_2)^2$ is the square of the total energy of the beams. One can see that this expression satisfies the necessary conditions: $s_e \cdot k_1 = 0$ and $s_e^2 = -1$. In the reaction (1) CMS this polarization four-vector can be written as $s_{e\mu} = (E/m)(s_0, \vec{s})$, where $s_0 = p/E, \vec{s} = \vec{p}/p$ and $\vec{p}(E)$ is the electron momentum (energy). So, we see that this vector describes the longitudinal polarization of the electron.

Then the spin–dependent part of the leptonic tensor $L^\gamma_{\mu\nu}(s_e)$ can be written as

$$L^\gamma_{\mu\nu}(s_e) = 2\frac{i}{s}(1 - 4\frac{m^2}{s})^{-1/2}\{B_1 < \mu\nu k_1 q > + B_2 < \mu\nu k_2 q >\}, \tag{35}$$

where the structure functions $B_i$ are

$$B_1 = -2\frac{s}{t_2}(1 - \frac{s}{t_1}) + 4m^2(\frac{1}{t_1} + \frac{1}{t_2})[1 - s(\frac{1}{t_1} + \frac{1}{t_2})] + 8m^4(\frac{1}{t_1} + \frac{1}{t_2})^2,$$

$$B_2 = 2\frac{s}{t_1}\left(1 - \frac{s}{t_2}\right) - 8m^2 \frac{q \cdot k}{t_1^2} + 8m^4\left(\frac{1}{t_1} + \frac{1}{t_2}\right)^2. \tag{36}$$

Let us consider the case of unpolarized initial beams and when final state is unpolarized or the final nucleon has polarization. Then the contraction of the spin–independent leptonic tensor and hadronic tensor which corresponds to the polarized nucleon can be written as

$$S^\gamma = L^\gamma_{\mu\nu}(0) H_{\mu\nu} = L^\gamma_{\mu\nu}(0)[H_{\mu\nu}(0) + H_{\mu\nu}(1)] = S^\gamma(0) + S^\gamma(1), \tag{37}$$

where $S^\gamma(0)(S^\gamma(1))$ corresponds to the unpolarized particles (polarized nucleon).

We can write for the unpolarized part

$$S^\gamma(0) = L^\gamma_{\mu\nu}(0)H_{\mu\nu}(0) = L^\gamma_{\mu\nu}(0)[\bar{H}_1 \tilde{g}_{\mu\nu} + \frac{\bar{H}_2}{M^2}\tilde{p}_{1\mu}\tilde{p}_{1\nu}] = \bar{H}_1 Q_1 + \frac{\bar{H}_2}{M^2}Q_2, \quad (38)$$

where the bar over the structure functions $H_i, i=1,2$ means that they depend on the shifted variable $q^2$ which is equal to $q^2 = s - t_1 - t_2$. In the reaction (1) CMS it is $q^2 = 4E(E-\omega)$, where $\omega$ is the hard photon energy. The kinematic coefficients $Q_{1,2}$ can be written as

$$Q_1 = -8 + \frac{4}{t_1 t_2}(s^2 + q^4 + 4m^2 s) - 8m^2(q^2 + 2m^2)\left(\frac{1}{t_1} + \frac{1}{t_2}\right)^2, \quad (39)$$

$$Q_2 = 2q^2\left(\frac{\chi_1}{t_2} + \frac{\chi_2}{t_1} + \frac{1}{t_1 t_2}[\chi_1^2 + \chi_2^2 - s(\chi_1 + \chi_2)]\right) + 4m^2 \frac{q^4}{t_1 t_2} +$$

$$+4m^2\left(\frac{\chi_1}{t_2} - \frac{\chi_2}{t_1}\right)\left(\frac{q^2 - \chi_1}{t_2} - \frac{q^2 - \chi_2}{t_1}\right) + 2M^2\left[2\frac{q^2 s}{t_1 t_2} + \frac{t_1}{t_2} + \frac{t_2}{t_1} - 2m^2 q^2\left(\frac{1}{t_1} + \frac{1}{t_2}\right)^2\right],$$

where $\chi_{1,2} = 2p_1 \cdot k_{1,2}$.

The contraction of the unpolarized lepton tensor and the hadron tensor corresponding to the polarized nucleon has the form

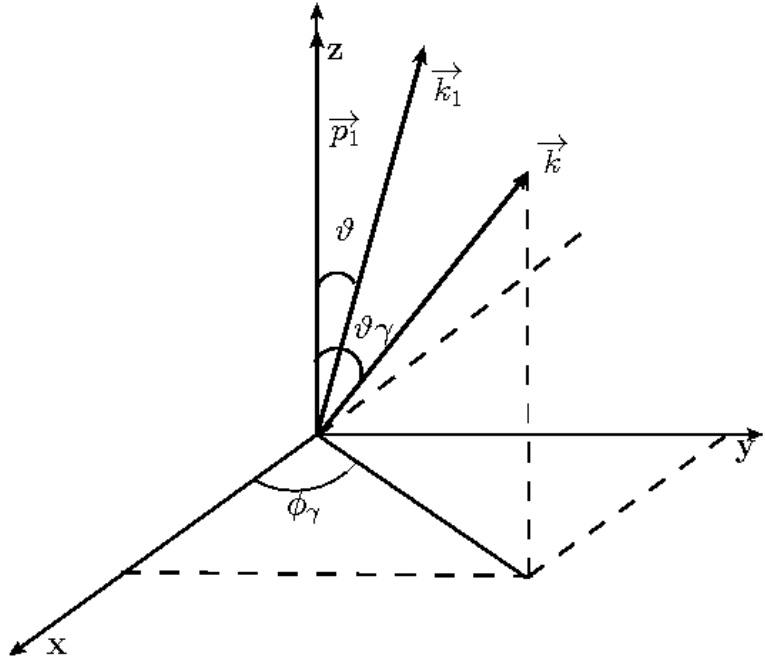

Fig. 1. The angles defining the kinematics of the reaction (1) in its CMS.

$$S^\gamma(1) = L^\gamma_{\mu\nu}(0)H_{\mu\nu}(1) = \frac{2}{M}(1-\tau)^{-1} Im G_M(q^2) G_E^*(q^2) Q_3, \quad (40)$$

where the function $Q_3$ can be written as

$$Q_3 = -4y_1\left[\frac{q^2}{t_1 t_2}(2\chi_1 + t_1 - s) + 2m^2\left(\frac{2\chi_2 - q^2}{t_1 t_2} - \frac{2\chi_1 - q^2}{t_2^2}\right)\right] - 4y_2\left[\frac{q^2}{t_1 t_2}(2\chi_2 + t_2 - s) + 2m^2\left(\frac{2\chi_1 - q^2}{t_1 t_2} - \frac{2\chi_2 - q^2}{t_1^2}\right)\right], (41)$$

$$y_1 = <k_1 p_1 p_2 s>, \quad y_2 = <k_2 p_1 p_2 s>.$$

Let us choose in the reaction CMS the following coordinate system: $z$ axis is directed along the nucleon momentum $\vec{p}_1$, the momentum of the initial electron beam $\vec{k}_1$ forms the $xz$ plane (the angle between these two momenta is $\vartheta$), $y$ axis is directed along the vector $\vec{k}_1 \times \vec{p}_1$ The momentum of the emitted photon $\vec{k}$ is defined by the polar and azimuthal angles, $\vartheta_\gamma$ and $\phi_\gamma$, respectively. The angles defining the kinematics of the reaction (1) in its CMS are given in Fig.1.

Then the cross section of the process (1) can be written as

$$d\sigma^\gamma = \frac{1}{32\pi^3} \frac{\alpha^3}{sq^4} L^\gamma_{\mu\nu} H_{\mu\nu} \frac{d^3\vec{k}}{\omega} \frac{d^3\vec{p}_1}{E_1} \frac{d^3\vec{p}_2}{E_2} \delta^{(4)}(k_1 + k_2 - p_1 - p_2 - k), \tag{42}$$

where $E_1(E_2)$ and $\omega$ are the energies of the nucleon (antinucleon) and photon, respectively; $\vec{p}_2$ is the antinucleon momentum.

On the basis of this expression we can obtain the different distributions depending on the experimental conditions. If we measure the nucleon scattering angle and variables of the emitted photon, we can obtain the following distribution

$$\frac{d\sigma^\gamma}{d\Omega d\Omega_\gamma d\omega} = \frac{1}{32\pi^3} \frac{\alpha^3}{sq^4} \omega |\vec{p}_1| [W - \omega + \omega \frac{E_1}{|\vec{p}_1|} \cos\vartheta_\gamma]^{-1} S^\gamma. \tag{43}$$

The dependence of this expression on the azimuthal angle $\phi_\gamma$ is due to the denominators $t_1$ and $t_2$. The integration over this variable can be done easily. In this case the quantity $q^2$ is $q^2 = W(W - 2\omega)$. Using the energy and momentum conservation: $W = \omega + E_1 + E_2$ and $\vec{k} + \vec{p}_1 + \vec{p}_2 = 0$, we can obtain the following relation between the nucleon energy $E_1$ and the photon polar angle $\vartheta_\gamma$

$$W(W - 2\omega) - 2(W - \omega)E_1 = 2\omega |\vec{p}_1| \cos\vartheta_\gamma. \tag{44}$$

If we retain the energies of the final hadrons we can obtain another distribution, namely

$$\frac{d\sigma^\gamma}{d\Omega dE_1 dE_2 d\phi_\gamma} = \frac{1}{32\pi^3} \frac{\alpha^3}{sq^4} S^\gamma. \tag{45}$$

In this case $q^2 = W[2(E_1 + E_2) - W]$. The integration over the photon azimuthal angle $\phi_\gamma$ can be done easily. Instead of antinucleon energy $E_2$ we can introduce the accolinearity angle $\chi$ (the angle between the momenta of the nucleon and antinucleon) which is measured in some experiments. From the energy and momentum conservation we have

$$2|\vec{p}_1||\vec{p}_2|\cos\chi = W^2 + 2M^2 - 2W(E_1 + E_2) + 2E_1 E_2. \tag{46}$$

If the additional photon is absent then $\chi = 180^0$ and the deviation of this angle from the value $180^0$ means that there is an additional photon. Experimenters choose the events in some interval of this angle. Using the above expression we can change $dE_2$ by $d\cos\chi$. We have

$$d\cos\chi = -\frac{dE_2}{2|\vec{p}_1||\vec{p}_2|^3}[2M^2(E_1 - W) + E_2(W^2 - 2WE_1 + 2M^2)]. \tag{47}$$

The energy of the antinucleon $E_2$ can be expressed in terms of the nucleon energy $E_1$ and the accolinearity angle $\chi$ by the following way

$$E_2 = \frac{(W - E_1)(W^2 - 2WE_1 + 2M^2) \pm |\vec{p}_1|\cos\chi\sqrt{W^2(W - 2E_1)^2 - 4M^2|\vec{p}_1|^2 \sin^2\chi}}{2[(W - E_1)^2 - |\vec{p}_1|^2 \cos^2\chi]}. \tag{48}$$

The polar angle of the emitted photon $\vartheta_\gamma$ can be also expressed in terms of the energies of the final hadrons. We have

$$\cos\vartheta_\gamma = \frac{(W-E_1)(E_1+2E_2-W)-E_1^2}{2|\vec{p}_1|(W-E_1-E_2)}. \tag{49}$$

Let us parameterize the nucleon spin four-vector $s_{1\mu}$ in terms of the four-momenta of the particles participating in the reaction under study. When measuring the polarization of the produced particle the $z$-axis is usually chosen along the momentum of this particle. So, in the reaction CMS we choose the longitudinal direction $\vec{l}$ ($z$ axis) along the nucleon momentum and the transverse one $\vec{t}$ in the plane $(\vec{p}_1, \vec{k}_1)$ and perpendicular to $\vec{l}$, and denote these polarization four-vectors as $P_\mu^{(i)}, i = l, t, n$

$$P_\mu^{(l)} = \frac{1}{d_l}[(\chi_1+\chi_2)p_{1\mu} - 2M^2(k_1+k_2)_\mu], d_l^2 = M^2[(\chi_1+\chi_2)^2 - 4M^2 s],$$

$$P_\mu^{(t)} = \frac{1}{d_t}\{(\chi_1+\chi_2)(\chi_2 k_{1\mu} - \chi_1 k_{2\mu}) + s[(\chi_1-\chi_2)p_{1\mu} - 2M^2(k_1-k_2)_\mu]\},$$

$$d_t^2 = s(\chi_1\chi_2 - M^2 s)[(\chi_1+\chi_2)^2 - 4M^2 s],$$

$$P_\mu^{(n)} = -\frac{1}{d_n}<\mu k_1 k_2 p_1>, d_n^2 = \frac{s}{4}(\chi_1\chi_2 - M^2 s). \tag{50}$$

It can be verified that the set of the four–vectors $P_\mu^{(l,t,n)}$ has the properties

$$P_\mu^{(\alpha)} P_\mu^{(\beta)} = -\delta_{\alpha\beta}, P_\mu^{(\alpha)} \cdot p_{1\mu} = 0, \alpha, \beta = l, t, n \tag{51}$$

and that in the reaction CMS we have

$$P_\mu^{(l)} = \frac{1}{M}(|\vec{p}_1|, E_1\vec{n}_1), P_{m\mu}^{(t)} = (0, \vec{t}), P_\mu^{(n)} = (0, \vec{n}), \tag{52}$$

$$\vec{n}_1 = \frac{\vec{p}_1}{|\vec{p}_1|}, \vec{n}_2 = \frac{\vec{k}_1}{|\vec{k}_1|}, \vec{t} = \frac{\vec{n}_2 - (\vec{n}_1 \cdot \vec{n}_2)\vec{n}_1}{\sqrt{1-(\vec{n}_1 \cdot \vec{n}_2)^2}}, \vec{n} = \frac{\vec{n}_2 \times \vec{n}_1}{\sqrt{1-(\vec{n}_1 \cdot \vec{n}_2)^2}}.$$

Note that in the Born case (there is no additional photon) these nucleon polarization 4-vectors coincides with the ones given by the Eq. (16).

The variables $y_{1,2}$ in Eq. (41) have the following expressions for particular choice of the nucleon polarization four–vectors $P_\mu^{(i)}, i = l, t, n$

$$y_2^l = -y_1^l = 2\frac{M^2}{d_l}<k_1 k_2 p_1 p_2>, \quad y_i^t = \frac{<k_1 k_2 p_1 p_2>}{d_t}[2M^2 s - \chi_i(\chi_1+\chi_2)], \quad i=1,2,$$

$$y_1^n = \frac{1}{8d_n}\{s[s\chi_1 - 2M^2(s-t_1)] + \chi_1[\chi_2(s-t_1) - \chi_1(s-t_2)]\}, \tag{53}$$

$$y_2^n = -\frac{1}{8d_n}\{s[s\chi_2 - 2M^2(s-t_2)] + \chi_2[\chi_1(s-t_2) - \chi_2(s-t_1)]\}.$$

The function $Q_3$ depends on the nucleon polarization four–vectors $P_\mu^{(i)}, i = l, t, n$ and it determines the spin–dependent part of the cross section. The function $Q_3$ has the following expressions for a particular choice of the nucleon

polarization four–vectors $P_\mu^{(l,t,n)}$ (in the reaction CMS)

$$Q_3^{(l)} = 4M s\omega \sin\vartheta \sin\vartheta_\gamma \sin\phi_\gamma [(E-\omega)\left(\frac{1}{t_1} - \frac{1}{t_2} + 4W|\vec{p}_1|\cos\vartheta \frac{1}{t_1 t_2}\right) -$$

$$-2m^2|\vec{p}_1|\cos\vartheta\left(\frac{1}{t_1^2} + \frac{1}{t_2^2}\right) - 2m^2(E_1 - E + \omega)\left(\frac{1}{t_1^2} - \frac{1}{t_2^2}\right)],$$

$$Q_3^{(t)} = 4s\omega \sin\vartheta_\gamma \sin\phi_\gamma \{(E-\omega)[(p_1 - E_1\cos\vartheta)\frac{1}{t_1} + (p_1 + E_1\cos\vartheta)\frac{1}{t_2} - \tag{54}$$

$$-8\frac{E}{t_1 t_2}p_1(E-\omega)(E_1\cos^2\vartheta + E - E_2) + 2\frac{m^2}{t_1^2}(p_1 + E_1\cos\vartheta)(p_1\cos\vartheta + E - E_2) + 2\frac{m^2}{t_2^2}(p_1 - E_1\cos\vartheta)(E - E_2 - p_1\cos\vartheta)\},$$

$$Q_3^{(n)} = 4\frac{E}{p_1}\frac{1}{\sin\vartheta}[L_0 + L_+\left(\frac{1}{t_1} + \frac{1}{t_2}\right) + L_-\left(\frac{1}{t_1} - \frac{1}{t_2}\right) + 2\frac{L_{12}}{t_1 t_2} + 2\overline{L}_+\left(\frac{1}{t_1^2} + \frac{1}{t_2^2}\right) + 2\overline{L}_-\left(\frac{1}{t_1^2} - \frac{1}{t_2^2}\right)],$$

where for the coefficients $L_i$ we have

$$L_0 = -4E_1 p_1(E-\omega)\cos\vartheta, \quad L_{12} = 2sq^2 p_1\cos\vartheta[E^2 + E_1(2E_1 - 3E) - p_1^2\sin^2\vartheta],$$

$$L_+ = 2q^2 p_1\cos\vartheta[E_1(5E - 2E_1) - E(E + E_2) + p_1^2\sin^2\vartheta],$$

$$L_- = q^2[2EE_1(\omega - E) + E_1^2(E_1 - E_2) + 2M^2 E_2 + p_1^2(2E_1 + 2E - \omega)\cos^2\vartheta],$$

$$\overline{L}_+ = -4m^2 E p_1\cos\vartheta[2E(E - E_2)(E + E_1 - E_2) - M^2\omega - (2E - \omega)p_1^2\sin^2\vartheta],$$

$$\overline{L}_- = -4m^2 E(E - E_2)[2EE_1(E + E_1 - E_2) - M^2(2E + \omega) - (4E - \omega)p_1^2\sin^2\vartheta].$$

Let us integrate the expressions $Q_3^{(i)}, i = l, t, n,$ over the angular variable $\phi_\gamma$. We have the following integrals

$$I_{1,2} = \int \sin\phi_\gamma \frac{d\phi_\gamma}{t_{1,2}}, \quad J_{1,2} = \int \sin\phi_\gamma \frac{d\phi_\gamma}{t_{1,2}^2}, \quad I_{12} = \int \sin\phi_\gamma \frac{d\phi_\gamma}{t_1 t_2}, \tag{55}$$

where

$$t_{1,2} = A_{1,2} + B_{1,2}\cos\phi_\gamma, A_{1,2} = 2\omega(E \mp k_1\cos\vartheta\cos\vartheta_\gamma), B_{1,2} = \pm 2\omega k_1\sin\vartheta\sin\vartheta_\gamma.$$

After integration we have

$$I_i = -\frac{1}{B_i}\ln(A_i + B_i\cos\phi_\gamma), \quad J_i = \frac{1}{B_i}\frac{1}{A_i + B_i\cos\phi_\gamma}, \quad i = 1, 2,$$

$$I_{12} = \frac{1}{2E\omega B_1}\ln\left|\frac{A_2 + B_2\cos\phi_\gamma}{A_1 + B_1\cos\phi_\gamma}\right|. \tag{56}$$

If we integrate over the whole possible region of the angle $\phi_\gamma$, i.e., over $(0, 2\pi)$, we have that all these integrals are equal to zero. So, in this case only perpendicular (to the reaction plane) polarization of the nucleon gives nonzero

contribution (as well as in the Born approximation). We have the following integrals

$$\int \frac{d\phi_\gamma}{t_1} = \frac{\pi}{\Delta_-}, \int \frac{d\phi_\gamma}{t_2} = \frac{\pi}{\Delta_+}, \Delta_\mp = \omega k_1 \left[ \left( \frac{E}{k_1} \cos\vartheta \mp \cos\vartheta_\gamma \right)^2 + \frac{m^2}{k_1^2} \sin^2\vartheta \right]^{1/2},$$

$$\int \frac{d\phi_\gamma}{t_1^2} = \frac{\pi}{4\Delta_-^3} A_1, \int \frac{d\phi_\gamma}{t_2^2} = \frac{\pi}{4\Delta_+^3} A_2, \int \frac{d\phi_\gamma}{t_1 t_2} = \frac{\pi}{2k \cdot q} \left( \frac{1}{\Delta_-} + \frac{1}{\Delta_+} \right). \tag{57}$$

So, we have after integration over the angle $\phi_\gamma$

$$Q^{(n)} = \int Q_3^{(n)} d\phi_\gamma = \frac{E}{p_1} \frac{4\pi}{\sin\vartheta} \left[ 2L_0 + L_+ \left( \frac{1}{\Delta_-} + \frac{1}{\Delta_+} \right) + L_- \left( \frac{1}{\Delta_-} - \frac{1}{\Delta_+} \right) + \frac{L_{12}}{k \cdot q} \left( \frac{1}{\Delta_-} + \frac{1}{\Delta_+} \right) + \right.$$

$$\left. + \frac{1}{2} \bar{L}_+ \left( \frac{A_1}{\Delta_-^3} + \frac{A_2}{\Delta_+^3} \right) + \frac{1}{2} \bar{L}_- \left( \frac{A_1}{\Delta_-^3} - \frac{A_2}{\Delta_+^3} \right) \right]. \tag{58}$$

## RADIATIVE RETURN. SMALL ANGLES

Since the main contribution, proportional to the large logarithm, comes from the integration of the integrand in the case of collinear kinematics of photon emission, we consider this case. For definiteness let us consider the case when the emitted photon moves close to the initial electron direction:

$$\widehat{k_1 k} = \vartheta \leq \vartheta_0 \ll 1, \vartheta_0 \gg \frac{m}{E}.$$

The differential cross section can be written as

$$\frac{d\sigma^\gamma}{d\Omega dx d\Omega_\gamma} = \frac{1}{32\pi^2} \frac{\alpha^3}{(1-x)^2 s^2} xy_1 [2 - x + x\cos\vartheta \frac{Y_1}{y_1}]^{-1} L_{\mu\nu}^\gamma H_{\mu\nu}, \tag{59}$$

where $x$ is the energy fraction carried away by the emitted photon, $x = \omega / E$ and $y_1 = p_1 / E$, $Y_1 = E_1 / E$ (here $p_1(E_1)$ is the momentum (energy) of the nucleon in the reaction CMS).

Integrating the leptonic tensor over the photon angular variables we obtain the following result for the case of unpolarized initial beams

$$L_{\mu\nu}^{\gamma(un)} = \int L_{\mu\nu}^\gamma(0) d\Omega_\gamma = -\frac{16\pi^2}{x\alpha E^2} F(L, x) [E^2 \tilde{g}_{\mu\nu} + (1-x)\tilde{k}_{1\mu}\tilde{k}_{1\nu}], \tag{60}$$

where the factor $F(L, x)$ has the form [36]

$$F(L, x) = \frac{\alpha}{2\pi} [(L_s - 1) \frac{1 + (1-x)^2}{x} + x + \frac{1 + (1-x)^2}{x} L_\theta], L_s = \ln \frac{s}{m^2}, L_\theta = \ln \frac{\vartheta_0^2}{4}. \tag{61}$$

After the integration over the photon variables the differential cross section can be written as

$$\frac{d\sigma^\gamma}{d\Omega dx} = \frac{\alpha^2}{s^3} \frac{y_1}{(1-x)^2} F(L, x) [2 - x + x\cos\vartheta \frac{Y_1}{y_1}]^{-1} S^{un}, \tag{62}$$

where we introduce

$$S^{un} = -[E^2 \tilde{g}_{\mu\nu} + (1-x)\tilde{k}_{1\mu}\tilde{k}_{1\nu}] H_{\mu\nu}.$$

Then the differential cross section for the case of the unpolarized particles can be written as

$$\frac{d\sigma^{\gamma}_{un}}{d\Omega dx} = \frac{\alpha^2}{2s}\frac{y_1}{(1-x)^2}F(L,x)[2-x+x\cos\vartheta\frac{Y_1}{y_1}]^{-1}D(x), \tag{63}$$

where the factor $D(x)$ is

$$D(x) = -\frac{1}{4}\frac{y^2}{1-x}\frac{1}{x+y^2-1}\{[4(1-x)(1-Y_1)(Y_1+x-1)-y_1^2(2-x)^2\sin^2\vartheta- \tag{64}$$

$$-x^2y^2](\tau|G_M(q^2)|^2-|G_E(q^2)|^2)-8(1-\tau)(1-x)^2|G_M(q^2)|^2\},$$

where $y = M/E$ and $q^2 = (1-x)s$, $\tau = q^2/4M^2$. The nucleon electromagnetic form factors $G_M$ and $G_E$ are taken at the value $q^2$. Thus, the emission of the photon decreases the argument of the form factors. If we remove the emission factor $F(L,x)dx$ and put $x=0$ we obtain the standard expression for the differential cross section in the Born approximation since $D(0) = D$. Setting the form factors equal to one, $G_E = G_M = 1$, we obtain the differential cross section for the case of the point particles in the final state, for example, $e^+e^- \to \mu^+\mu^-$ or $\tau^+\tau^-$. To do this, it is necessary to replace $D(x)$ by $D^{point}(x)$ where

$$D^{point}(x) = \frac{1}{4}\frac{1}{1-x}\{4(1-x)[(1-x)(1+Y_1)+Y_1(1-Y_1)]-y_1^2(2-x)^2\sin^2\vartheta-x^2y^2\}. \tag{65}$$

The nucleon energy $E_1$, the scattering angle $\vartheta$ and $x$ variable are connected by the following relation

$$xy_1\cos\vartheta = 2(1-x)-(2-x)Y_1.$$

Using this relation we can determine the nucleon energy as a function of two variables: $x$ and $\cos\vartheta$. We have

$$Y_1 = [(2-x)^2 - x^2\cos^2\vartheta]^{-1}\{2(1-x)(2-x) \pm x\cos\vartheta[4(1-x)^2 - y^2((2-x)^2 - x^2\cos^2\vartheta)]^{1/2}\}. \tag{66}$$

When calculating the radiative corrections to the polarization observables it is convenient to parameterize the nucleon polarization 4–vector in terms of the four–momenta of the particles participating in the reaction under study. Any four-vector $U_\mu^{(i)}$ which parameterize the polarization state of the particle can be read as

$$U_\rho^{(i)} = U_\rho^{(i)}(k_1,k_2,p_1,p_2). \tag{67}$$

Let us imagine for a moment that chosen parametrization on the right side of Eq. (67) is stabilized relative substitution $k_1 \to xk_1$

$$U_\mu^{(i)}(k_1,k_2,p_1,p_2) = U_\mu^{(i)}(xk_1,k_2,p_1,p_2). \tag{68}$$

Further we call such parametrization a stable one (relative to the substitution $k_1 \to xk_1$) and denote such set of polarization four-vectors as $S_\mu^{(i)}, i = l,t,n$.

In what follows we use the following set of the stable polarization four-vectors

$$S_\mu^{(l)} = \frac{1}{M\chi_1}(2M^2k_{1\mu}-\chi_1 p_{1\mu}), \quad S_\mu^{(n)} = \frac{1}{n_1}<\mu k_2 k_1 p_1>, \quad n_1^2 = \frac{s}{4}(\chi_1\chi_2-M^2s), \tag{69}$$

$$S_\mu^{(t)} = \frac{1}{n_2}[\chi_1 k_{2\mu}-sp_{1\mu}-(\chi_2-2M^2\frac{s}{\chi_1})k_{1\mu}], \quad n_2^2 = s(\chi_1\chi_2-M^2s).$$

If the polarization four-vectors $U_\mu^{(i)}$ are unstable ones under above substitution they can be always expressed in terms of the stable polarization four-vectors by means of some linear combination

$$U^{(i)}_\mu(k_1,k_2,p_1,p_2) = A_{ij}(k_1,k_2,p_1,p_2)U^{(j)}_\mu(k_1,k_2,p_1,p_2). \tag{70}$$

In case of the radiative return at small angles we have only one plane (reaction plane) and, therefore, normal polarization four-vector $S^{(n)}_\mu$ do not mix with the polarization four-vectors $P^{(l,t)}_\mu$. In this case we have some rotation in the reaction plane and it can be written down as

$$S^{(l)}_\mu = cos\psi P^{(l)}_\mu + sin\psi P^{(t)}_\mu, \quad S^{(t)}_\mu = cos\psi P^{(t)}_\mu - sin\psi P^{(l)}_\mu, \tag{71}$$

where

$$cos\psi = \frac{E_1 cos\theta - p_1}{E_1 - p_1 cos\theta}, \quad sin\psi = -\frac{M sin\theta}{E_1 - p_1 cos\theta}. \tag{72}$$

Since in the time–like region the nucleon form factors are complex functions, the $H_{\mu\nu}(1)$ tensor has a symmetric part (in $\mu,\nu$ indices) and, therefore, the contraction of the $L^{\gamma(un)}_{\mu\nu}$ and $H_{\mu\nu}(1)$ tensors is not zero. This leads to the nucleon polarization in the case when the rest of the particles are unpolarized. The polarization vector in this case is normal to the reaction plane and can be written as

$$D(x)P_y = 2yy_1 \frac{sin\vartheta}{x+y^2-1}(Y_1 - 1 + x + y_1 cos\vartheta)ImG_E(q^2)G^*_M(q^2). \tag{73}$$

Putting $x = 0$ we obtain the result of the Born approximation. For the point final particles we have natural result that the polarization is equal to zero.

Let us consider the case when the electron beam is longitudinally polarized. Integrating the leptonic tensor $L^\gamma_{\mu\nu}(s_e)$ over the photon angular variables we obtain the following result [36]

$$L^{\gamma(pol)}_{\mu\nu} = \int L^\gamma_{\mu\nu}(s_e)d\Omega_\gamma = i\frac{8\pi^2}{x\alpha E^2}F^{pol}(L,x)<\mu\nu k_1 k_2>, \tag{74}$$

where the factor $F^{pol}(L,x)$, describing the photon emission by the longitudinally polarized electron beam, has the form

$$F^{pol}(L,x) = \frac{\alpha}{2\pi}[(L_s - 1)\frac{1+(1-x)^2}{x} - x + \frac{1+(1-x)^2}{x}L_\theta]. \tag{75}$$

The contraction of the tensors $L^{\gamma(pol)}_{\mu\nu}$ and $H_{\mu\nu}(0)$ is equal to zero and, therefore, the asymmetry determined by the longitudinal polarization of the electron beam is absent. A non–zero asymmetry may arise due to the two–photon exchange contribution, for example. The contraction of the $L^{\gamma(pol)}_{\mu\nu}$ and antisymmetrical part (in $\mu,\nu$ indices) of the $H_{\mu\nu}(1)$ tensors is non zero. This leads to the nucleon polarization and in this case the polarization vector lies in the reaction plane. The components of the nucleon polarization vector for the case of the stable set of the polarization four-vectors can be written down as (we assume 100% polarization of the electron beam)

$$F(L,x)D(x)P_x = 2sin\vartheta \frac{F^{pol}(L,x)}{Y_1 - y_1 cos\vartheta}\frac{yy_1}{x+y^2-1}\{(1-x)(1-Y_1+y_1 cos\vartheta)|G_M(q^2)|^2 - \tag{76}$$

$$-[y^2 + (1-x)(y_1 cos\vartheta - Y_1)]ReG_E(q^2)G^*_M(q^2)\},$$

$$F(L,x)D(x)P_z = -\frac{1}{2}\frac{F^{pol}(L,x)}{Y_1 - y_1 cos\vartheta}\frac{1}{x+y^2-1}\{(1-x)(1-Y_1+y_1 cos\vartheta)[y^2 + (1-x)(y_1 cos\vartheta - \tag{77}$$

$$-Y_1)]|G_M(q^2)|^2 + y^2[(1-x)(Y_1 - y_1 cos\vartheta)(2 - Y_1 + y_1 cos\vartheta) - y^2]ReG_E(q^2)G^*_M(q^2)\}.$$

If we remove the unpolarized and polarized emission factors $F(L,x)dx$ and $F^{pol}(L,x)dx$ and put $x = 0$ we obtain the standard expressions for the components of the nucleon polarization vector in the Born approximation. For the case of the point final particles we have

$$F(L,x)D(x)P_x^{point} = -y(2-x)sin\vartheta F^{pol}(L,x), \qquad (78)$$

$$F(L,x)D(x)P_z^{point} = [xy_1 + (2-x)Y_1 cos\vartheta]F^{pol}(L,x).$$

Let us consider the case when the produced antinucleon and nucleon are both polarized. Then the components of the polarization correlation tensor $P_{ik}, (i,k=x,y,z)$, of the nucleon and antinucleon (determined as a coefficient in front of $\xi_{1i}\xi_{2k}$ where $\vec{\xi}_1(\vec{\xi}_2)$ is the nucleon (antinucleon) polarization vector in its rest frame) can be written as

$$D(x)P_{yy} = \frac{z}{1-x-y^2}[(1-x)|G_M(q^2)|^2 - y^2|G_E(q^2)|^2], \qquad (79)$$

$$D(x)P_{xx} = (2-2x+x^2)sin^2\vartheta|G_M(q^2)|^2 + z(1+dx^2sin^2\vartheta)\frac{1}{\tau-1}(\tau|G_M(q^2)|^2 - |G_E(q^2)|^2) -$$

$$-dsin^2\vartheta\{x[4(1-x)+(2-x)(xy-2Y_1)-2y_1 cos\vartheta]|G_M(q^2)|^2 - \frac{x}{\tau-1}[-4(1-x)^2+2(1-x)(y+Y_1) +$$

$$+xy(Y_1-y)](\tau|G_M(q^2)|^2 - ReG_E(q^2)G_M^*(q^2)) + \frac{x^2}{2y^2}\frac{z}{(\tau-1)^2}[y(y-Y_1)-2(1-x)]|G_E(q^2)-G_M(q^2)|^2\},$$

$$D(x)P_{zz} = \frac{1}{y^2}\{2yY_1 cos^2\vartheta - 2y_1(y_1+xcos\vartheta)+x(2-x)[y(y_1-Y_1 cos\vartheta)cos\vartheta+(y_1+xcos\vartheta)(y_1-$$

$$-Y_1 cos\vartheta)]\}|G_M(q^2)|^2 + \frac{z}{y^2}\frac{1}{\tau-1}\{y_1(y_1+xcos\vartheta)+yY_1[1+d(y_1+xcos\vartheta)^2]\}(\tau|G_M(q^2)|^2 - |G_E(q^2)|^2) -$$

$$-\frac{1}{y^2}\frac{1}{\tau-1}\{x(2-x)[yy_1-Y_1(y_1+xcos\vartheta)]cos\vartheta - y_1(y_1+xcos\vartheta)(2-x)^2 + x^2yY_1 cos^2\vartheta\}$$

$$[\tau|G_M(q^2)|^2 - ReG_E(q^2)G_M^*(q^2) + \frac{1}{2y^2}\frac{z}{\tau-1}|G_E(q^2)-G_M(q^2)|^2] +$$

$$+\frac{d}{y}(y_1+xcos\vartheta)(y_1 cos\vartheta+x)\{[2Y_1 cos\vartheta+x(2-x)(y_1-Y_1 cos\vartheta)]|G_M(q^2)|^2 -$$

$$-\frac{x}{2y^2}\frac{z}{(\tau-1)^2}[(2-x)y_1+xY_1 cos\vartheta]|G_E(q^2)-G_M(q^2)|^2 -$$

$$-\frac{1}{(\tau-1)}[(2-x^2)y_1+x^2Y_1 cos\vartheta - 2(1-x)y_1(Y_1-y_1 cos\vartheta)](\tau|G_M(q^2)|^2 - ReG_E(q^2)G_M^*(q^2))\} -$$

$$-\frac{2}{y^2}\frac{1}{(\tau-1)}(1-x)(1-Y_1+y_1 cos\vartheta)cos\vartheta[yy_1+Y_1(y_1+xcos\vartheta)](\tau|G_M(q^2)|^2 - ReG_E(q^2)G_M^*(q^2)),$$

$$D(x)P_{xz} = -sin\vartheta\{(2-2x+x^2)cos\vartheta|G_M(q^2)|^2 + \frac{1}{y}(y_1+xcos\vartheta)[-x(2-x)+yd(y_1 cos\vartheta+x)(2-$$

$$-2x+x^2)]|G_M(q^2)|^2 + \frac{1}{y}\frac{1}{(\tau-1)}(y_1+x\cos\vartheta)[xyzd(\tau|G_M(q^2)|^2-|G_E(q^2)|^2)-$$

$$-2(1-x)(1+y_1\cos\vartheta-Y_1)(\tau|G_M(q^2)|^2-\mathrm{Re}G_E(q^2)G_M^*(q^2))]\}-\frac{x}{y}\frac{1}{(\tau-1)}\sin\vartheta\{(y_1+x\cos\vartheta)[2-x-$$

$$-xyd(y_1\cos\vartheta+x)]-xy\cos\vartheta\}[\frac{z}{2y^2}\frac{1}{(\tau-1)}|G_E(q^2)-G_M(q^2)|^2+\tau|G_M(q^2)|^2-\mathrm{Re}G_E(q^2)G_M^*(q^2)],$$

$$D(x)P_{zx}=-\frac{\sin\vartheta}{y}\{(2-2x+x^2)Y_1\cos\vartheta+xd[2(1-x)y_1(y-\sin^2\vartheta Y_1)+y(2-x)(yy_1-y_1Y_1-$$

$$-xY_1\cos\vartheta)]\}|G_M(q^2)|^2-\frac{x}{y}zd\frac{\sin\vartheta}{(\tau-1)}[y_1(2+y)+x(Y_1\cos\vartheta-y_1)](\tau|G_M(q^2)|^2-|G_E(q^2)|^2)-$$

$$-x\frac{zd}{2y^3}\frac{\sin\vartheta}{(\tau-1)^2}[y_1(2-x)+xY_1\cos\vartheta][2(1-x)+y(Y_1-y)]|G_M(q^2)-G_E(q^2)|^2-$$

$$-d\frac{x}{y}\frac{\sin\vartheta}{\tau-1}\{-2y\cos\vartheta(Y_1-y)(Y_1+y_1\cos\vartheta)+2y_1(1-x)(2Y_1+2y+y_1\cos\vartheta)+x^2yy_1+$$

$$+Y_1\cos\vartheta[8(1-x)-Y_1-4y_1\cos\vartheta+(2-x^2)y]+\frac{1}{xd}[(2-4x+x^2)y_1-2y_1(Y_1+y_1\cos\vartheta)-$$

$$-x^2Y_1\cos\vartheta]\}(\tau|G_M(q^2)|^2-\mathrm{Re}G_E(q^2)G_M^*(q^2)),$$

where we introduce the notations

$$z=y^2+(1-x)(Y_1-y_1\cos\vartheta)(Y_1-y_1\cos\vartheta-2), \quad d=\frac{1}{y(y+Y_2)}, \quad Y_2=\frac{E_2}{E}.$$

For completeness we give also the nonzero coefficients in the case of longitudinally polarized electron beam

$$F(L,x)D(x)P_{xy}=-\frac{1}{2}xy_1\frac{\sin^2\vartheta}{\tau-1}F^{pol}(L,x)\mathrm{Im}G_M(q^2)G_E^*(q^2),$$

$$F(L,x)D(x)P_{yx}=-\frac{1}{2}xy_1d\frac{\sin^2\vartheta}{\tau-1}F^{pol}(L,x)[2(1-x)+y(Y_1-y)]\mathrm{Im}G_M(q^2)G_E^*(q^2), \qquad (80)$$

$$F(L,x)D(x)P_{zy}=\frac{1}{2}\frac{y_1}{y}\frac{\sin\vartheta}{\tau-1}F^{pol}(L,x)[(2-x)y_1+xY_1\cos\vartheta]\mathrm{Im}G_M(q^2)G_E^*(q^2),$$

$$F(L,x)D(x)P_{yz}=\frac{y_1}{2}\frac{\sin\vartheta}{\tau-1}F^{pol}(L,x)\{d(y_1+x\cos\vartheta)[(2-x)y+2(1-x)]-x\cos\vartheta\}\mathrm{Im}G_M(q^2)G_E^*(q^2).$$

.
### PHOTON ENERGY DISTRIBUTION

Let us calculate the differential cross section of the reaction (1) for the experimental conditions when only the energy of the collinear photon is measured. To do this, it is necessary to calculate the quantity $d\sigma^\gamma/dx$, i.e., we have to integrate the differential cross section (62) over the nucleon angular variables. The invariant integration of the hadronic tensor is the simplest method to do this.

Let us define the following quantity

$$F_{\mu\nu} = \int H_{\mu\nu} \frac{d^3\vec{p}_1}{(2\pi)^3 2E_1} \frac{d^3\vec{p}_2}{(2\pi)^3 2E_2} \delta^{(4)}(k_1 + k_2 - p_1 - p_2 - k). \tag{81}$$

Using the requirements of the Lorentz invariance and $q_\mu H_{\mu\nu} = q_\nu H_{\mu\nu} = 0$ (it is a consequence of the hadron current conservation) the most general expression for the $F_{\mu\nu}$ tensor can be written as

$$F_{\mu\nu} = A(q^2)\tilde{g}_{\mu\nu}. \tag{82}$$

Only unpolarized part of the hadronic tensor $H_{\mu\nu}(0)$ gives nonzero contribution to the Eq. (81) since we integrate over the variables of the nucleon–antinucleon pair. To calculate the $A(q^2)$ function it is necessary to multiply by $g_{\mu\nu}$ tensor the left and right sides of Eq. (81) and this leads to the following result

$$A(q^2) = \frac{1}{3}\int H_{\mu\nu}(0) g_{\mu\nu} \frac{d^3\vec{p}_1}{(2\pi)^3 2E_1} \frac{d^3\vec{p}_2}{(2\pi)^3 2E_2} \delta^{(4)}(k_1 + k_2 - p_1 - p_2 - k). \tag{83}$$

The contraction of the hadron $H_{\mu\nu}(0)$ and $g_{\mu\nu}$ tensors can be written in terms of the structure functions $H_i(q^2), i = 1, 2$

$$H_{\mu\nu}(0) g_{\mu\nu} = 3H_1(q^2) + M^2(1-\tau)H_2(q^2). \tag{84}$$

This contraction can be written also in terms of the nucleon electromagnetic form factors

$$H_{\mu\nu}(0) g_{\mu\nu} = -8M^2[|G_E(q^2)|^2 + 2\tau |G_M(q^2)|^2]. \tag{85}$$

The easiest way to do the integration over the variables of the nucleon–antinucleon final state is to use the center–of–mass system of this pair. As a result we have

$$A(q^2) = -\frac{2}{3}\frac{M^2}{(2\pi)^5}\sqrt{1 - \frac{4M^2}{q^2}}[|G_E(q^2)|^2 + 2\tau |G_M(q^2)|^2]. \tag{86}$$

Representing the square of the matrix element as contraction of the lepton and hadron tensors we can write the expression for the differential cross section as (we integrate over the angular variables of the collinear photon)

$$\frac{d\sigma^\gamma}{dx} = 4\pi^4 x \alpha^3 \frac{1}{q^4} F_{\mu\nu} L_{\mu\nu}^{\gamma\,un}. \tag{87}$$

Since the $F_{\mu\nu}$ tensor is symmetric (over the $\mu, \nu$ indices) only unpolarized part of the lepton tensor gives nonzero contribution. As a result we have for the differential cross section $d\sigma^\gamma / dx$ (where $x$ is the beam energy fraction carried away by the emitted collinear (to the direction of the electron momentum) photon in the $e^+e^- \to N\bar{N}\gamma$ reaction)

$$\frac{d\sigma^\gamma}{dx} = \frac{8\pi}{3}\frac{\alpha^2}{q^4} M^2 \sqrt{1 - \frac{4M^2}{q^2}} F(L,x)[|G_E(q^2)|^2 + 2\tau |G_M(q^2)|^2]. \tag{88}$$

The limit of the point–like final particles (for example, the $e^+e^- \to \mu^+\mu^-\gamma$ reaction) is given by

$$\frac{d\sigma^\gamma_{point}}{dx} = \frac{4\pi}{3}\frac{\alpha^2}{q^2}\sqrt{1 - \frac{4M^2}{q^2}}(1 + 2\frac{M^2}{q^2})F(L,x). \tag{89}$$

Let us write the energy spectrum of the collinear photon in the terms of the structure functions, which define the spin–independent part of the hadron tensor, for the case of two final particles with different masses. We have

$$\frac{d\sigma^\gamma}{dx} = \frac{\pi}{3}\frac{\alpha^2}{q^4}\sqrt{(1+\frac{M_1^2-M_2^2}{q^2})^2 - 4\frac{M_1^2}{q^2}}F(L,x)\{-3H_1(q^2) + \frac{1}{4M_1^2 q^2}[(q^2+M_1^2-M_2^2)^2 - 4M_1^2 q^2]H_2(q^2)\}. \quad (90)$$

And the spin–independent part of the hadron tensor is written in this case as

$$H_{\mu\nu}(0) = H_1(q^2)\tilde{g}_{\mu\nu} + \frac{1}{M_1^2}H_2(q^2)\tilde{p}_{1\mu}\tilde{p}_{1\nu}, \quad (91)$$

where $M_1$ is the mass of the particle with four–momentum $p_{1\mu}$. Using this equation we can calculate the contributions of some various channels to the energy spectrum.

Spin–zero particles.

Consider the production of two spinless particles (for example, $e^+e^- \to \pi^+\pi^-\gamma, K^+K^-\gamma$). The electromagnetic current of the hadrons in this case is determined by one complex form factor and the explicit expression of this current can be written as

$$J_\mu = F(q^2)(p_1 - p_2)_\mu \phi_1^* \phi_2^*, \quad (92)$$

where $p_{1\mu}(p_{2\mu})$ and $\phi_1(\phi_2)$ are the four–momentum and wave function of the first (second) spinless particle. The structure functions in this case are

$$H_1(q^2) = 0, H_2(q^2) = M^2 |F(q^2)|^2, \quad (93)$$

where $M$ is the spinless particle mass.

Spin–one particles.

Consider the production of two spin–one particles (for example, $e^+e^- \to \rho^+\rho^-\gamma, d\bar{d}\gamma$). The electromagnetic current for a spin–one particle is completely described by three complex form factors. Assuming the $P$- and $C$-invariance of the hadron electromagnetic interaction, this current can be written as

$$J_\mu = (p_1 - p_2)_\mu [-G_1(q^2)U_1^* \cdot U_2^* + \frac{1}{M^2}G_3(q^2)(U_1^* \cdot q U_2^* \cdot q - \frac{q^2}{2}U_1^* \cdot U_2^*] - G_2(q^2)(U_{1\mu}^* U_2^* \cdot q - U_{2\mu}^* U_1^* \cdot q), \quad (94)$$

where $U_{1\mu}(U_{2\mu})$ is the polarization four–vector describing the first (second) spin–one particle.

The form factors $G_i(q^2)$ are related to the standard electromagnetic form factors: $G_C$ (charge monopole), $G_M$ (magnetic dipole) and $G_Q$ (charge quadrupole) by

$$G_M = -G_2, G_Q = G_1 + G_2 + 2G_3, G_C = -\frac{2}{3}\tau(G_2 - G_3) + (1 - \frac{2}{3}\tau)G_1. \quad (95)$$

The standard electromagnetic form factors have the following normalizations

$$G_C(0) = 1, G_M(0) = \mu, G_Q(0) = -M^2 Q,$$

where $\mu(Q)$ is the magnetic (quadrupole) moment of a spin–one particle. The structure functions in this case are

$$H_1(q^2) = 8M^2\tau(1-\tau)|G_M|^2, H_2(q^2) = 12M^2(|G_C|^2 - \frac{2}{3}\tau|G_M|^2 + \frac{8}{9}\tau^2|G_Q|^2), \quad (96)$$

where $M$ is the spin–one particle mass. The form factors of the $\rho$–meson, deuteron in the time–like region of the momentum transfer were discussed in Refs. [37, 38].

Channel $\pi a_1$.

Let us consider the production of $\pi$–meson and $a_1(1260)$, where $a_1(1260)$ is the axial–vector meson with the following quantum numbers $I^G(J^{PC}) = 1^-(1^{++})$. This channel gives a substantial contribution to the $e^+e^- \to 4\pi$ process. In the energy region $1 \leq W \leq 2.5$ GeV ($W$ is the total energy of the colliding beams) the process of four pion production is one of the dominant processes of the reaction $e^+e^- \to$ hadrons.

The electromagnetic current of the $\gamma^* \to \pi a_1$ transition is described by two complex form factors. Assuming the

$P$ - and $C$ - invariance of the hadron electromagnetic interaction this current can be written as [39]

$$J_\mu = F_1(q^2)(q^2 U^*_\mu - q\cdot U^* q_\mu) + F_2(q^2)(q\cdot p_2 U^*_\mu - q\cdot U^* p_{2\mu}), \tag{97}$$

where $U_\mu$ is the polarization four–vector describing the spin–one $a_1$–meson, $p_{2\mu}$ is the pion four–momentum and $F_i(q^2)$ are the electromagnetic form factors describing the $\gamma^* \to \pi a_1$ transition. The structure functions corresponding to this current are

$$H_1(q^2) = -|q^2 F_1 + \frac{1}{2}(q^2 - M_1^2 + M_2^2)F_2|^2, \quad H_2(q^2) = q^2[q^2|F_1+F_2|^2 - M_1^2|F_2|^2], \tag{98}$$

where $M_1(M_2)$ is the $a_1$–meson (pion) mass.

Channel $\Delta\bar{N}$.

Let us consider the production of antinucleon and $\Delta(1232)$–isobar with the $I(J^P)=3/2(3/2^+)$. One can expect that this channel gives a substantial contribution to the $e^+e^- \to N\bar{N}\pi$ process. The electromagnetic current of the $\gamma^* \to \Delta\bar{N}$ transition is described by three complex form factors and it can be written as [40]

$$J_\mu = \bar{u}_\alpha(p_1)\gamma_5\{F_1(q^2)q_\alpha[q^2\gamma_\mu + (M_1+M_2)q_\mu] + \tag{99}$$

$$+F_2(q^2)q_\alpha[\gamma_\mu + \frac{1}{M_1-M_2}(p_1-p_2)_\mu] + F_3(q^2)[(M_1+M_2)g_{\mu\alpha} + q_\alpha\gamma_\mu]\}v(p_2),$$

where $M_1(M_2)$ and $p_1(p_2)$ are the mass and four–momentum of the $\Delta$–isobar (antinucleon). $F_i(q^2)(i=1,2,3)$ are the electromagnetic form factors describing the $\gamma^* \to \Delta\bar{N}$ transition. The structure functions corresponding to this current are

$$H_1(q^2) = -\frac{1}{3}(M_1 M_2 + r)[\frac{1}{M_1^2}(M_1 M_2 - r)|F|^2 + (M_1+M_2)^2|F_3|^2 - \frac{1}{M_1}(M_1 M_2 - r)(M_1+M_2)ReFF_3^*], \tag{100}$$

$$H_2(q^2) = \frac{2}{3}(M_1^2 M_2^2 - r^2)[|F|^2 - 2ReFF_3^*] - \frac{1}{3}(M_1+M_2)[(M_1-M_2)(M_2^2-M_1^2) +$$

$$+M_2 q^2]ReFF_3^* - \frac{1}{3}M_1^2(M_1 M_2 + r)[\frac{4q^2}{(M_1-M_2)^2}|F_2|^2 - \frac{1}{M_1^2}|(M_1+M_2)F_3 + \frac{q^2+M_1^2-M_2^2}{M_1-M_2}F_2|^2],$$

where $F = q^2 F_1 + F_2 + F_3$, $r = (q^2 - M_1^2 - M_2^2)/2$.

## RADIATIVE RETURN. LARGE ANGLES

Let us consider the experimental conditions when the nucleon–antinucleon pair is not detected. We assume that the apparatus detect events in the whole phase space of the nucleon–antinucleon pair. So, it is necessary to integrate the differential cross section (42) over the variables of this pair. This procedure is already done in the Section "PHOTON ENERGY DISTRIBUTION". Using these results we have for the energy and angular distribution of the photon

$$\frac{d\sigma^\gamma}{dx d\Omega_\gamma} = 4\pi^4 x \alpha^3 \frac{1}{q^4} A(q^2)\tilde{g}_{\mu\nu} L^\gamma_{\mu\nu}(0), \tag{101}$$

where the function $A(q^2)$ has the following form for the general case of two particles with unequal masses

$$A(q^2) = \frac{1}{12}\frac{1}{(2\pi)^5}\frac{1}{q^2}\sqrt{(q^2+M_1^2-M_2^2)^2 - 4M_1^2 q^2} \tag{102}$$

$$\{3H_1(q^2) - \frac{1}{4M_1^2 q^2}[(q^2 + M_1^2 - M_2^2)^2 - 4M_1^2 q^2]H_2(q^2)\},$$

where $M_1$ and $M_2$ are the masses of the final particles. The structure functions $H_i(q^2)$ describe the hadronic tensor $H_{\mu\nu}(0)$ for the case of two–particle final state. The expressions of the structure functions in terms of the electromagnetic form factors of the transition $\gamma^* \to h_1 h_2$ can be found in the Section "PHOTON ENERGY DISTRIBUTION" for some particular final states. After calculation of the convolution of the $L_{\mu\nu}^\gamma(0)$ and $\tilde{g}_{\mu\nu}$ tensors we have for the distribution (neglecting terms proportional to the electron mass)

$$\frac{d\sigma^\gamma}{dxd\Omega_\gamma} = 32\pi^4 x\alpha^3 \frac{1}{q^4} A(q^2)\{1 - \frac{1}{t_1 t_2}[q^4 + 2q^2 q \cdot k + 2(q \cdot k)^2]\}. \tag{103}$$

Let us choose the $z$ axis along the electron momentum, and the photon momentum in the $xz$ plane. Then integrating over the whole range of the photon azimuthal angle we obtain

$$\frac{d\sigma^\gamma}{dxdc} = 2(2\pi)^5 x\alpha^3 \frac{1}{q^4} A(q^2)[1 - \frac{2}{x^2}\frac{2 - 2x + x^2}{1 - \beta_e^2 c^2}], \tag{104}$$

where $\beta_e^2 = 1 - 4m^2/s$, $c = \cos\vartheta_\gamma$, $\vartheta_\gamma$ is the angle between the electron and photon momenta.

Integrating this expression over the photon polar angle within the range $m/E \ll \vartheta_1 < \vartheta_\gamma < \vartheta_2$, where $(\pi - \vartheta_2) \gg m/E$, we obtain the following expression for the differential cross section

$$\frac{d\sigma^\gamma}{dq^2} = 2(2\pi)^5 \alpha^3 \frac{1}{sq^4} A(q^2)[(c_1 - c_2)(1 - \frac{q^2}{s}) - \frac{1 + q^4/s^2}{1 - q^2/s}\ln\frac{(1+c_1)(1-c_2)}{(1-c_1)(1+c_2)}], \tag{105}$$

where $c_i = \cos\vartheta_i$. Let us rewrite this formula in another form introducing the total cross section of the $e^+e^- \to h_1 h_2$ reaction. We have

$$\frac{d\sigma^\gamma}{dq^2} = \frac{2}{3}\frac{\alpha^3}{sq^2} R(q^2)[(c_1 - c_2)(1 - \frac{q^2}{s}) - \frac{1 + q^4/s^2}{1 - q^2/s}\ln\frac{(1+c_1)(1-c_2)}{(1-c_1)(1+c_2)}], \tag{106}$$

where we define

$$R(q^2) = \frac{\sigma(e^+e^- \to h_1 h_2)}{\sigma(e^+e^- \to \mu^+\mu^-)} = \frac{1}{4}\frac{1}{q^4}\sqrt{(q^2 + M_1^2 - M_2^2)^2 - 4M_1^2 q^2} \times \tag{107}$$

$$\{-3H_1(q^2) + \frac{1}{4M_1^2 q^2}[(q^2 + M_1^2 - M_2^2)^2 - 4M_1^2 q^2]H_2(q^2)\}.$$

If we integrate the expression (104) over the photon polar angle within the following range $m/E \ll \vartheta_\gamma^{min} < \vartheta_\gamma < \pi - \vartheta_\gamma^{min}$ we obtain the following result for the differential cross section of the $e^+e^- \to h_1 h_2 \gamma$ reaction

$$\frac{d\sigma^\gamma}{dq^2} = \frac{4}{3}\frac{\alpha^3}{sq^2} R(q^2)[\frac{1 + q^4/s^2}{1 - q^2/s}\ln\frac{1 + c_m}{1 - c_m} - (1 - \frac{q^2}{s})c_m], \tag{108}$$

where $c_m = \cos\vartheta_\gamma^{min}$. If we put $h_1 h_2 = N\bar{N}$, then $R(q^2) = (\beta/2)[2|G_{MN}(q^2)|^2 + (1/\tau)|G_{EN}(q^2)|^2]$ and the expression for $d\sigma^\gamma/dq^2$ for the case $h_1 h_2 = N\bar{N}$ coincides with the one obtained in Ref. [30].

**CONCLUSIONS**

The emission of the hard photon by the electron or positron is considered in the $e^- + e^+ \to N + \bar{N} + \gamma$. The nucleon polarization and the differential cross sections for some experimental conditions have been calculated.

The case of the emission of the collinear (with respect to the direction of the electron beam momentum) photon is considered separately. The differential cross section, the nucleon polarization, the correlation coefficients for both polarized nucleon and antinucleon (provided the electron beam is unpolarized or longitudinally polarized), the transfer polarization from the longitudinally polarized electron beam to the nucleon have been calculated.

We have calculated the photon energy distribution for the reaction $e^+ e^- \to B_1 \bar{B}_2 \gamma$, where $B_1 (\bar{B}_2)$ is some baryon (antibaryon) for the case of the collinear (with respect to the direction of the electron beam momentum) photon, emitted in the initial state. As $B_1 \bar{B}_2$ final state we considered some channels, namely: two spinless mesons (for example, $\pi^+ \pi^-, K^+ K^-$), two spin–one particles (for example, $\rho^+ \rho^-, d\bar{d}$), the $\pi a_1(1260)$ and $\Delta(1232)\bar{N}$ channels. The photon energy distributions have been calculated in terms of the electromagnetic form factors of the $\gamma^* \to B_1 \bar{B}_2$ transition.

**Acknowledgments**

This work was partially supported by the Ministry of Education and Science of Ukraine (Projects No. 0115U000474 and No. 0117U004866). The research was conducted in the scope of the IDEATE International Associated Laboratory (LIA).